\documentclass[singlecolumn,amssymb, nobibnotes, show---pacs, superscriptaddress, aps, prd]{revtex4}
\usepackage{graphicx,amsmath,amssymb,color}

\usepackage{soul}

\begin{document}
\title{Magneto-optical rotation: Accurate approximated analytical solutions for single-probe atomic magnetometers}

\author{L. Deng$^*$}
\affiliation{Center for Optics Research and Engineering (CORE), Shandong University (Qingdao), China}
\email{*lu.deng@email.sdu.edu.cn}
\author{Claire Deng}
\affiliation{Thomas Wootton HS, Rockville, Maryland, USA 20850}

\date{\today}

\begin{abstract}
	We report an approximated analytical solution for a single-probe four-state atomic magnetometer where no analytical solution exists. This approximated analytical solution demonstrates excellent accuracy in broad probe power and detuning ranges when compared with the numerical solution obtained using a 4th order Runge-Kutta differential equation solver on MATLAB. The theoretical framework and results also encompass widely applied single-probe three-state atomic magnetometers for which no analytical solution, even approximated, is available to date in small detuning regions.  
\end{abstract}

\maketitle
	
	\noindent Magneto-optical Faraday rotation describes the rotation of the polarization plane of a linearly polarized light field traversing through a magnetized medium \cite{r1}. When a laser was first used as the light source, it was found that the angle of polarization plane rotation was dependent upon the intensity of the laser, hence given the term ``nonlinear magneto-optical rotation effect" (NMORE) \cite{r2}. For the past 60 years, atomic NMORE studies have primarily focused on a single probe laser interacting with $F=1$ atomic systems \cite{r3,r4,r5}, Many pioneering studies and innovations \cite{r6,r7,r8,r9,r10,r11,r12,r13,r14} have contributed to the advancement in this research field. However, to date there has been no analytic solution$-$or even an approximated analytic solution with sufficient accuracy$-$for NMORE phenomena. This is especially the case when the laser is tuned close to the relevant one-photon resonance in order to enhance the observability of NMORE signal \cite{r7}. Such a near resonance excitation inevitably introduces significant power broadening, rendering analytic solutions very difficult. Consequently, only numerical solutions are available for such near resonance excitation operations \cite{r3}. Recent experimental and theoretical progress \cite{r15a,r15}, especially the demonstration of an NMORE blockade in a single-probe configuration, has for the first time opened the possibility for analytical or approximated analytical solutions with high accuracy for single-probe NMOREs.
	
	\vskip 10pt
	\noindent In this work we show a high-accuracy approximate analytical solution for widely studied single-probe atomic systems. The theoretical framework presented here encompasses both three and four-state systems for both near and far-detuned optical fields [see Fig. 1(a)]. 
	
	\begin{figure}[htb]
		\centering
		\includegraphics[width=12 cm]{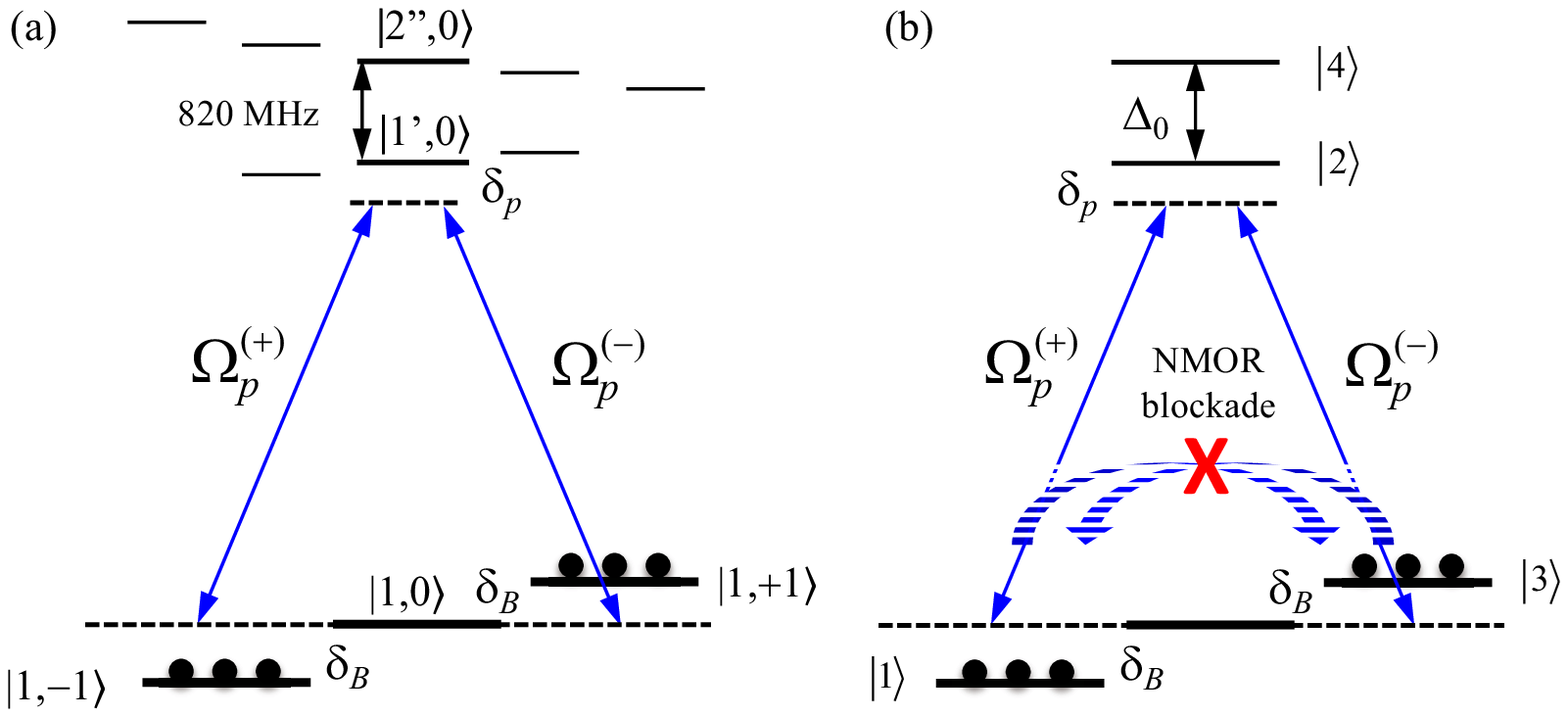}
		\caption{(a) Rubidium D-1 transitions relevant to single-probe atomic magnetometers. (b) A four-state model single-probe atomic magnetometer. Dashed blue arrows represent two opposite two-photon transitions. Population symmetry as well as transition symmetry lead to a NMORE blockade that strongly suppresses NMORE signal.}
	\end{figure}
	
	\vskip 10pt
	\noindent We consider a four-state atomic system, depicted in Fig. 1(b), where the atomic state $|j\rangle$ has energy $\hbar\omega_j$ ($j=1,...,4$) and the lower three states form an $F=1$ system. We assume that the probe field $\mathbf{E}_{p}$ (frequency $\omega_{p}$) is polarized along the $\hat{x}$-axis and propagates along the $\hat{z}$-axis. Its $\sigma^{(\pm)}$ components independently couple the $|1\rangle\Leftrightarrow|2\rangle$ and $|3\rangle\Leftrightarrow|2\rangle$ transitions with a large one-photon detuning $\delta_2=\omega_{p}-[\omega_2-(\omega_1+\omega_3)/2]=\delta_p$. The second excited state $|4\rangle$ is assumed to be $\Delta_0$ above the state $|2\rangle$, enabling the second excitation channel by the same probe field with a detuning of $\delta_4=\omega_{p}-[\omega_4-(\omega_1+\omega_3)/2]=\delta_{p}+\Delta_0$ (in our notation $\delta_{p}<0$ and $\Delta_0=-\omega_4+\omega_2<0$). This atomic system describes, for example the Rubidium D-1 line with a hyper-fine splitting, i.e., $|2\rangle=|5P_{1/2}, F'=1\rangle$ and $|4\rangle=|5P_{1/2}, F'=2\rangle$ with $|\Delta_0|=820$ MHz [Fig. 1(a)]. Since initially the population is equally shared by the two ground states $|F=1, m_F=\pm 1\rangle$, therefore two opposite two-photon transitions between states $|1\rangle$ and $|3\rangle$ with a two-photon detuning $2\delta_B$ are simultaneously established. Here, the circular components of the probe field simultaneously access excited states $|2\rangle$ and $|4\rangle$ with different excitation rates. The magnetic field induced Zeeman frequency shift $\delta_B=g\mu_0 B$ in the axial magnetic field $B=B_z$ is defined with respect to the mid-point between the two equally but oppositely shifted Zeeman levels $|1\rangle=|m_F=-1\rangle$ and $|3\rangle=|m_F=+1\rangle$. 
	
	\vskip 10pt
	\noindent Under the electric-dipole approximation, the system interaction Hamiltonian reads
	\begin{eqnarray}
	\frac{\hat{H}}{\hbar}\!=\!\sum_{j=1}^4\delta_j|j\rangle\langle j|\!+\!\sum_{m=2,4}\!\left[\Omega_{m1}|m\rangle\langle 1|\!+\!\Omega_{m3}|m\rangle\langle 3|\!+\!{\rm c.c}\right],\:
	\end{eqnarray}
	where $\delta_j$ is the laser detuning from state $|j\rangle$. The total electric field is given by $\mathbf{E}\!=\!\left(\mathbf{\hat{e}}_{+}{\cal E}_{p}^{(+)}\!+\!\mathbf{\hat{e}}_{-}{\cal E}_{p}^{(-)}\right){\rm e}^{i\theta_{p}}\!+\!{\rm c.c.}$ where $\theta_{p}\!=\!\mathbf{k}_{p}\!\cdot\mathbf{r}-\omega_{p}t$ with $k_{p}\!=\omega_{p}/c$ being the wavevector of the field. Expressing probe Rabi frequencies with respect to the state $|2\rangle$ as $\Omega_{21}\!=\!\Omega_{p}^{(+)}\!=\!D_{21}{\cal E}_{p}^{(+)}/2\hbar$ and $\Omega_{23}\!=\!\Omega_{p}^{(-)}\!=\!D_{23}{\cal E}_{p}^{(-)}/2\hbar$, we then have $\Omega_{41}\!=\!D_{41}{\cal E}_{p}^{(+)}/(2\hbar)\!=\!d_{42}\Omega_{p}^{(+)}$ and $\Omega_{43}\!=\!D_{43}{\cal E}_{p}^{(-)}/(2\hbar)\!=\!d_{42}\Omega_{p}^{(-)}$. Here, we define $d_{42}=D_{41}/D_{21}=D_{43}/D_{23}$ with $D_{nm}=\langle n|\hat{D}|m\rangle$ being the transition matrix element of the dipole operator $\hat{D}$ and $D_{21}=D_{23}$, $D_{41}=D_{43}$. 
	
	\vskip 10pt
	\noindent
	Under rotating wave approximation, the Schrodinger equations describing wave-function amplitudes are,
	\begin{subequations}
		\begin{align}
		&\dot{A}_{1}-i\delta_BA_1=i{\Omega_{p}^{(+)}}^*\left(A_2+d_{42}A_4\right)-\gamma A_1,\\
		&\dot{A}_{3}+i\delta_BA_3=i{\Omega_{p}^{(-)}}^*\left(A_2+d_{42}A_4\right)-\gamma A_3,\\
		&\dot{A}_{2}-i\delta_{p}A_2=i\Omega_{p}^{(+)}A_1+i\Omega_{p}^{(-)}A_3-\Gamma A_2,\\
		&\dot{A}_{4}\!-\!i\!\left(\delta_{p}\!-\!\Delta_0\right)\!A_4\!=\!id_{42}\!\left(\Omega_{p}^{(+)}A_1\!+\!\Omega_{p}^{(-)}A_3\right)\!-\!\Gamma\!A_4,
		\end{align}
	\end{subequations}
	where for simplicity we have expressed the ground and excited states' decay rates as $\gamma$ and $\Gamma$, respectively. 
	
	\vskip 10pt
	\noindent Applying the slowly varying envelope approximation and the third-order perturbation calculation \cite{r15,r16}, we obtain the Maxwell equations describing the evolution of both circular polarized components of the probe field $\mathbf{E}_{p}$ in the moving frame ($\xi=z-ct$, $\eta=z$), 
	
	\begin{equation}
	\frac{\partial\Omega_{p}^{(\pm)}}{\partial\eta}\approx-\sum_{n=2,4}\alpha_n\Omega_{p}^{(\pm)}\!+\!\sum_{n=2,4}\!\frac{\alpha_{n}}{2}(1+id_{n})S_{\mp}\Omega_{p}^{(\pm)}\!\!\left(\!\frac{\Gamma_{0}\!+\!i\beta_{0}}{\Gamma_{-}\!+\!i\beta_{-}}\!+\!\frac{\Gamma_{0}\!-\!i\beta_{0}}{\Gamma_{+}\!+\!i\beta_{+}}\!\right),
	\end{equation}
	where $\alpha_{n}=\kappa_{n}/\Gamma(1+d_{n}^2)$, $\kappa_{2}=\kappa_{23}$ and $\kappa_{4}=\kappa_{43}$ ($\kappa_2=2\pi|D_{21}|^2{\cal N}_0\omega_p/(\hbar c)$ where ${\cal N}_0$ is the atom number density). The first summation on the right accounts for the linear absorption (we have neglected the linear propagation phase shift since it does not contribute to NMORE). The normalized detunings are defined as 
	$d_{2}=\delta_{p}/\Gamma$, $d_{4}=(\delta_{p}-\Delta_0)/\Gamma$, and 
	$d_{B}=\delta_B/\gamma$. In addition, we have defined
	$\beta_{\pm}=-d_B\mp\beta_0S^{(\pm)}$, $\Gamma_{\pm}=1+\Gamma_0S^{(\pm)}$, $\beta_0=d_{2}/(1+d_2^2)+d_4/(1+d_4^2)$, and $\Gamma_0=1/(1+d_2^2)+1/(1+d_4^2)$ where $\beta_0$ and $\Gamma_0$ are the total light induced frequency shift and resonance broadening, respectively. The probe two-photon saturation parameters are defined as $S_{\pm}=|\Omega_{p}^{(\pm)}|^2/\gamma\Gamma$. 
	\vskip 10pt
	\noindent
	Letting $\Omega_{p}^{(\pm)}=R_{\pm}\,e^{i\theta_{\pm}}$ where $R_{\pm}$ and $\theta_{\pm}$ are real quantities, then Eq. (3) gives
	\begin{subequations}
		\begin{align}
		&\frac{\partial S_{\pm}}{\partial\eta}=-\alpha S_{\pm}\!+\!S_{+}S_{-}{\cal P}_{\pm},\\
		&\frac{\partial\theta_{\pm}}{\partial\eta}
		=\frac{S_{\mp}}{2}{\cal Q}_{\pm}.
		\end{align}
	\end{subequations}
	where ${\cal P}_{\pm}=\alpha{\cal A}\mp\alpha_{d}{\cal B}$, ${\cal Q}_{\pm}=\mp\alpha{\cal B}+\alpha_d{\cal A}$ with ${\cal A}\!=\!\Gamma_0(\Gamma_{-}/G_{-}\!+\!\Gamma_{+}/G_{+}\!)$, ${\cal B}\!=\!\Gamma_0(\beta_{-}/G_{-}\!+\!\beta_{+}/G_{+}\!)$, $\alpha=\alpha_2+\alpha_4$, $\alpha_d=d_2\alpha_2+d_4\alpha_4$, and $G_{\pm}=\Gamma_{\pm}^2+\beta_{\pm}^2$ (in the following calculations and without the loss of generality we neglect $\beta_0$ terms \cite{r17}). The advantage of this photon number representation is that intensity [i.e., Eq. (4a)] decouples from the phase [i.e., Eq. (4b)]. While both Eqs. (4a) and (4b) are highly nonlinear and complex, approximated analytical solutions can be obtained with excellent accuracy. As we show below, two key steps separately based on the physics of the single-probe system and mathematical considerations for approximation accuracy are necessary to achieve this.
	
	\vskip 10pt
	\noindent The first key step is to realize the presence of a symmetry-enforced NMORE blockade in any single-probe system where population and transition symmetry are present \cite{r18a}.
	We note that the probe field is the only energy source and therefore its two circular components must add up at any propagation distance to give $S_{-}(\eta)=S_0e^{-\alpha\eta}-S_{+}(\eta)$  where $S_0=\Omega_{p}(0)^2/\gamma\Gamma$ is the initial total energy of the single probe field. Taking the differential equation for the $S_{+}$ component in Eq. (4a) and inserting this energy restriction relation into intensity product in the second term on the right side of Eq. (4a), we immediately conclude a gain-clamping effect, i.e., $(S_0e^{-\alpha\eta}-S_{+})S_{+}$. This energy constraint locks the two probe components by enacting a self-restricted growth, resulting in $S_{\pm}(\eta)\approx S_{\pm}(0)$. That is, no appreciable magnetic field induced optical field change is allowed for either component (see numerical results and discussion later) \cite{r18}. It is this propagation growth restriction that limits the single-probe NMORE to be in characteristically {\bf linear}. This is a single-probe NMORE blockade first postulated in Ref.\cite{r15a} and then demonstrated mathematically in Ref.\cite{r15}. As we show below, it is this NMORE blockade and the linear absorption characteristics of both field components that lead to a high-accuracy approximate analytical solution that is well-applicable to both three and four-state atomic systems.
	
	\vskip 10pt
	\noindent The second key step is based on the mathematics consideration of ${\cal P}_{\pm}$ and ${\cal Q}_{\pm}$ which contain $S_{\pm}(\eta)$. One of the consequences of the above described NMORE blockade is that the dominant propagation behavior is determined by linear absorption since the gain is clamped. The magnetic field is contained in functions ${\cal P}_{\pm}$ and ${\cal Q}_{\pm}$, and its effect should mostly be near magnetic resonance. Therefore, it is a reasonable expectation that errors by replacing $S_{\pm}(\eta)$ in ${\cal P}_{\pm}$ and ${\cal Q}_{\pm}$ with $S_{\pm}(\eta)\rightarrow(S_0/2)e^{-\alpha\eta}$ should be relatively small. Therefore, as the first trial we set $S_{\pm}(\eta)=(S_{0}/2)e^{-\alpha\eta}$ in denominators of ${\cal P}_{\pm}$ and ${\cal Q}_{\pm}$.
	
	\vskip 10pt
	\noindent The major benefit of the second step is that now both ${\cal P}_{\pm}$ and ${\cal Q}_{\pm}$ contain $e^{-\alpha\eta}$ only, permitting analytical solutions to both field and NMORE. We get 
	\begin{equation}
	S_{\pm}(\eta; d_B)=\frac{S_0e^{S_0\Gamma_0{\cal L}_{\pm}}}{1+S_0e^{S_0\Gamma_0{\cal L}_{\pm}}}e^{-\alpha\eta},
	\end{equation}
	where
	\begin{equation*}
	{\cal L}_{\pm}(\eta; d_B)=2\Gamma_0\int_{0}^{\eta}\frac{\alpha(1+\Gamma_0S_0e^{-\alpha\eta}/2)\mp\alpha_dd_B}{d_B^2+(1+\Gamma_0S_0e^{-\alpha\eta}/2)^2}d\eta,
	\end{equation*}
	is analytically integrable. Inserting Eq. (5) into Eq. (4b), we obtain
	\begin{equation*}
	\theta_{\pm}(\eta; d_B)\!=\!\Gamma_0\!\int_{0}^{\eta}\!\!S_{\mp}(\eta; d_B)\frac{\alpha_d(1\!+\!\Gamma_0S_0e^{-\alpha\eta}/2)\!\pm\!\alpha d_B}{d_B^2\!+\!(1\!+\!\Gamma_0S_0e^{-\alpha\eta}/2)^2}d\eta.
	\end{equation*}
	Using symbolic evaluation routines on MATLAB or MATHEMATICA we obtain an analytical expression for the NMORE angle 
	\begin{subequations}
		\begin{align}
		\Theta(\eta; d_B)_{_{\rm NMORE}}=&-C_0\left(\frac{8}{15}\right)\frac{e^{-\alpha\eta}\Gamma_0S_0}{2\alpha(1+d_B^2)}\left\{i\left(\frac{\alpha+\alpha_dd_B}{1+{\cal M}_3{\cal M}_4^{i\frac{2\alpha_d}{\alpha}}}+\frac{\alpha-\alpha_dd_B}{1+{\cal M}_3{\cal M}_4^{-i\frac{2\alpha_d}{\alpha}}}\right){\rm ln}{\cal M}_1\right.\nonumber\\
		&\left.-\left(\frac{\alpha d_B-\alpha_d}{1+{\cal M}_3{\cal M}_4^{i\frac{2\alpha_d}{\alpha}}}+\frac{\alpha d_B+\alpha_d}{1+{\cal M}_3{\cal M}_4^{-i\frac{2\alpha_d}{\alpha}}}\right){\rm ln}{\cal M}_2\right\},
		\end{align}
	\end{subequations}where the amplitude adjustment constant $C_0$ is very close to unity and has a narrow range (typically $<\pm 5\%$) depending on the choices of relative transition strength, laser power and detuning (see numerical results below). In addition,
	\begin{subequations}
		\begin{align}
		{\cal M}_1&\!=\!\left[\frac{1\!+\!d_B^2\!+\!(1\!+\!id_B)\frac{\Gamma_0S_0}{2}}{1\!+\!d_B^2\!+\!(1\!-\!id_B)\frac{\Gamma_0S_0}{2}}\right]\!\left[ \frac{1\!+\!d_B^2\!+\!(1\!-\!id_B)e^{-\alpha\eta}\frac{\Gamma_0S_0}{2}}{1\!+\!d_B^2\!+\!(1\!+\!id_B)e^{-\alpha\eta}\frac{\Gamma_0S_0}{2}}\right],\\
		{\cal M}_2&=\left[\frac{d_B^2+(1+\frac{\Gamma_0S_0}{2})^2}{d_B^2+(1+e^{-\alpha\eta}\frac{\Gamma_0S_0}{2})^2}\right]e^{-2\alpha\eta},\\
		{\cal M}_3&=\frac{\left[d_B+i(1+\frac{\Gamma_0S_0}{2})\right]\left[d_B-i(1+e^{-\alpha\eta}\frac{\Gamma_0S_0}{2})\right]}{\left[d_B-i(1+\frac{\Gamma_0S_0}{2})\right]\left[d_B+i(1+e^{-\alpha\eta}\frac{\Gamma_0S_0}{2})\right]},\\
		{\cal M}_4&=\left[\frac{d_B^2+(1+e^{-\alpha\eta}\frac{\Gamma_0S_0}{2})^2}{d_B^2+(1+\frac{\Gamma_0S_0}{2})^2}\right]^2.
		\end{align}
	\end{subequations}
	\begin{figure}[htb]
		\centering
		\includegraphics[width=14 cm]{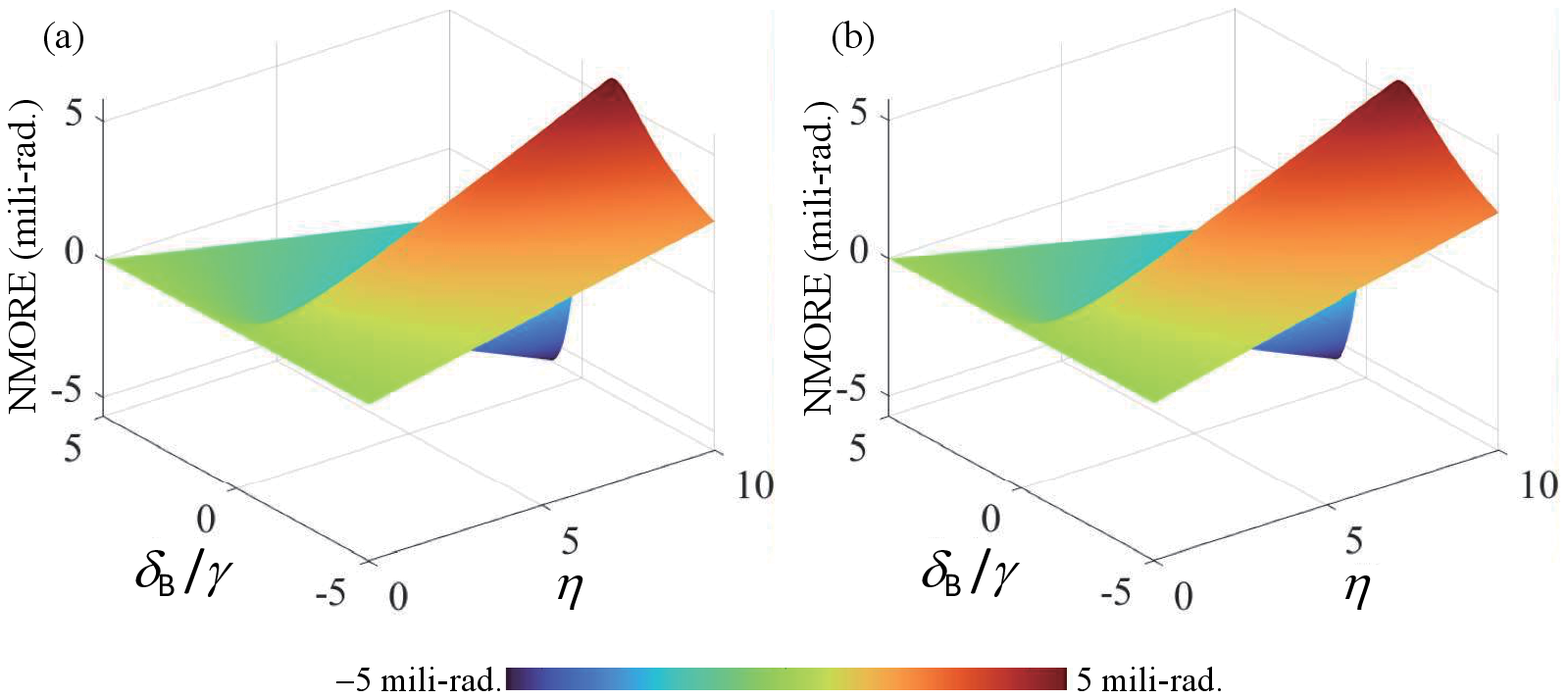}
		\caption{(a) Numerical solution by solving Eqs. (4a) and (4b) simultaneously. (b) NMORE evaluated using Eq. (6) by MATLAB. Parameters: $S_0=5$, $d_2=-5$, $\Delta_0=2$, $\kappa_4/\kappa_2=2$ ($\kappa_2=10^5$/cm.s), and $C_0=1.06$.  }
	\end{figure}
	\vskip 10pt
	\noindent 
	When the power broadening terms $\left(\frac{\Gamma_0S_0}{2}\right)$ in ${\cal M}_m$ ($m=1,..,4$) are neglected, we have ${\cal M}_1\rightarrow 1$, ${\cal M}_2\rightarrow e^{-2\alpha\eta}$, ${\cal M}_3\rightarrow 1$, and ${\cal M}_4\rightarrow 1$. Consequently, Eq. (6) gives
	\begin{equation}
	\Theta(\eta; d_B)_{_{\rm NMORE}}\approx-\frac{\alpha d_B}{2(1+d_B^2)}\Gamma_0S_0e^{-\alpha\eta}\eta,
	\end{equation}
	which is the NMORE for a four-level single-probe AM with power broadening neglected but linear absorption included \cite{r15}. When the state $|4\rangle$ is neglected, $\alpha=\alpha_2$,  $\Gamma_0=1/(1+d_2^2)$, and we recover the NMORE of a three-level single-probe AM without power broadening. Equations (5) and (6) are the most accurate approximated analytical solution to NMORE of three and four-state systems to date. The validity and accuracy of Eqs. (5) and (6) have been thoroughly verified using a 4th order Runge-Kutta numerical differential equation solver on MATLAB, as well as a MATHEMATICA symbolic evaluation routine for broad probe power and detuning ranges \cite{r19}.
	
	\vskip 10pt
	\noindent Figure 2(a) shows the NMORE of a four-state single-probe system obtained by a 4th order Runge-Kutta code that simultaneously solves Eqs. (4a) and (4b).  Figure 2(b) shows the NMORE evaluated using Eq. (6) with the same parameters. The small difference validates the proposal of replacing $S_{\pm}(\eta)\rightarrow(S_{0}/2)e^{-\alpha\eta}$ in ${\cal P}_{\pm}$ and ${\cal Q}_{\pm}$. When the numerical and the approximated analytical solutions are plotted at $\eta=10$ as a function of $d_B$ we have found that with $C_0=1.05\pm 0.02$ excellent agreement between the two methods in a broad probe detuning and power regions [also see Fig. 3], a testimonial of the accuracy of Eq. (6). 
	
	\begin{figure}[htb]
		\centering
		\includegraphics[width=14 cm]{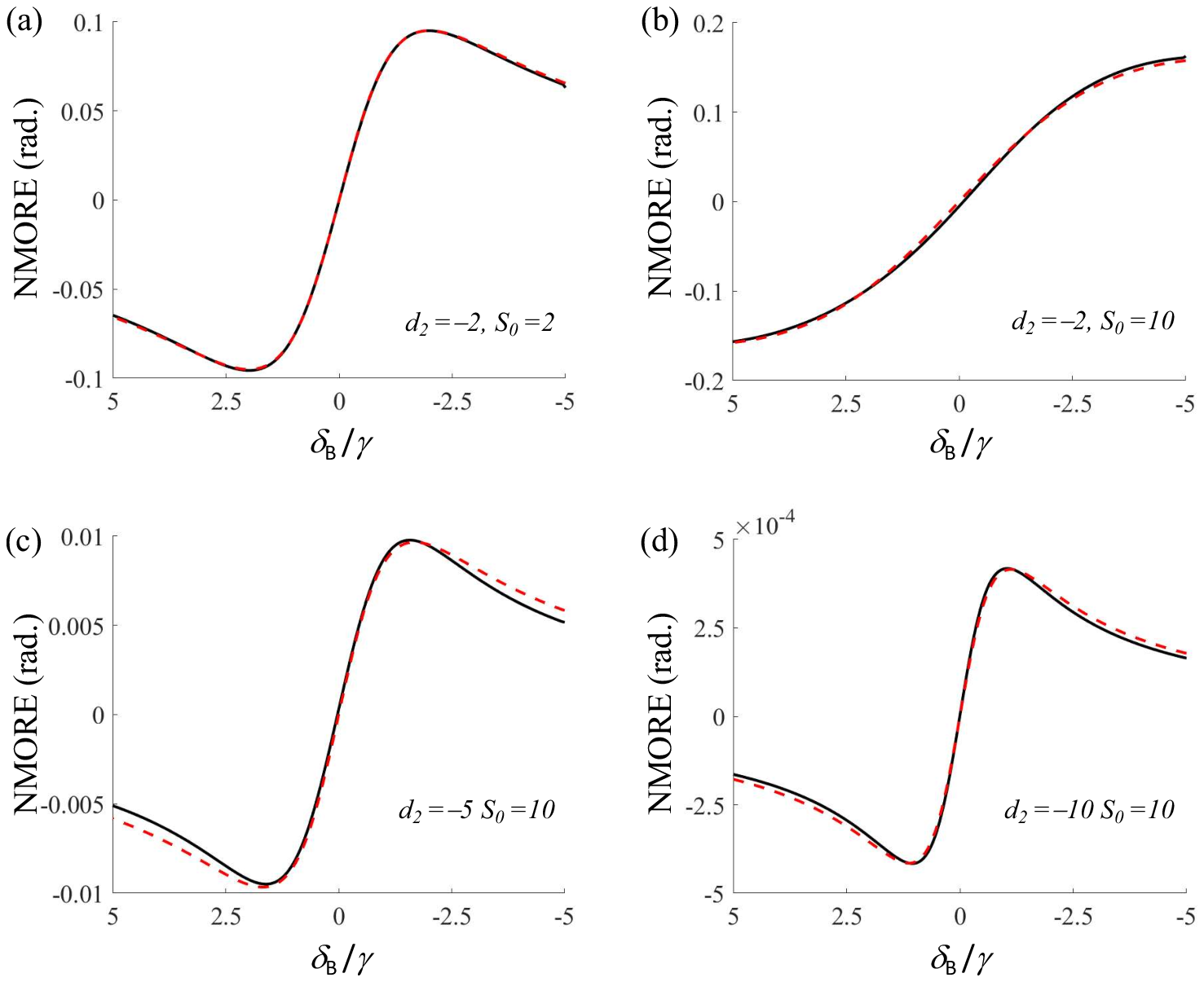}
		\caption{NMORE line profiles at $\eta=10$ by Eqs. (4a) and (4b) (black) and Eq. (6) (red). Figs. 3(a) and 3(b): $d_2=-2$ and probe powers are $S_0$=2 and 10, respectively. Here, a single amplitude adjustment constant $C_0=1.17$ is used for Eq. (6). Figs. 3(c) and 3(d): $S_0$=10 and probe detunings are $-5$ and $-10$, respectively, with $C_0=1.15\pm 0.02$ is used in Eq. (6).}
	\end{figure}
	
	\begin{figure}[htb]
		\centering
		\includegraphics[width=14 cm]{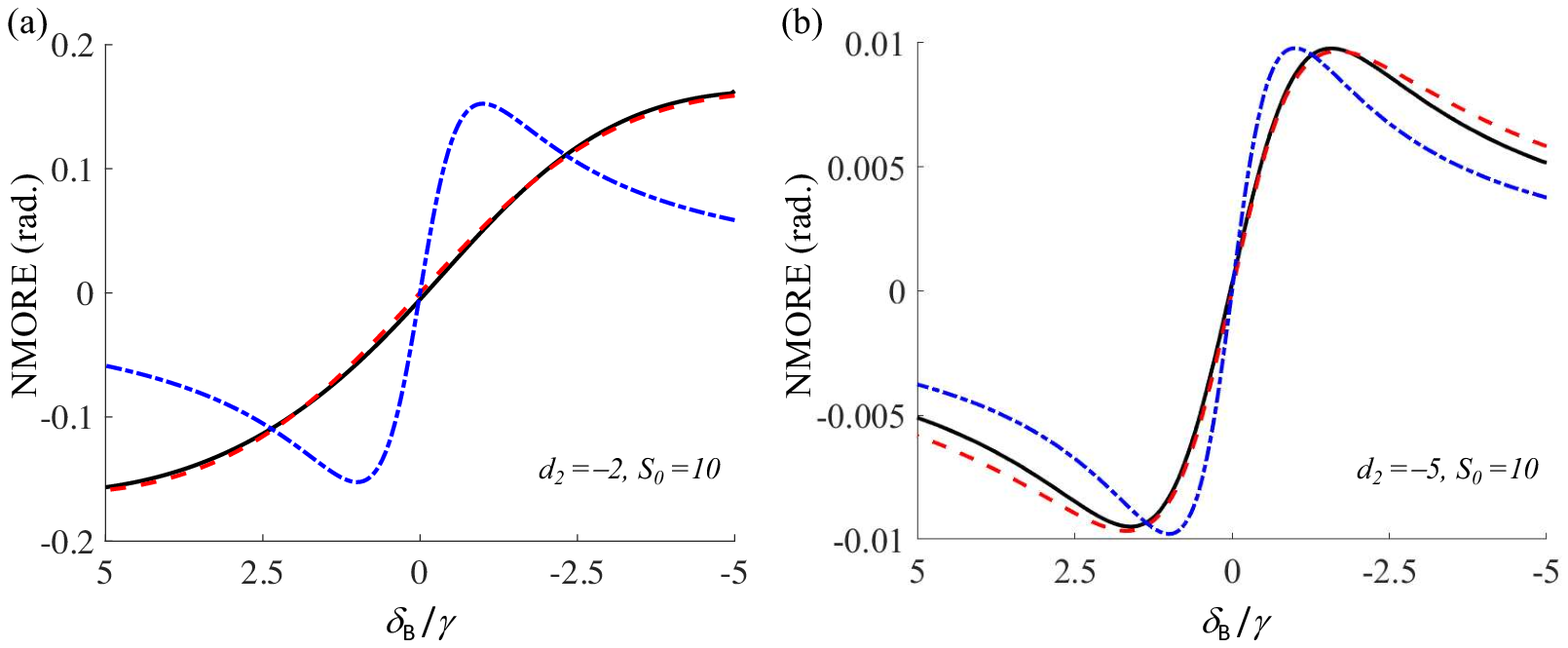}
		\caption{NMORE profiles obtained by numerical solution of Eqs. (4a) and (4b) (black), solutions from Eq. (6) (red) and Eq. (8) (blue) using $S_0=10$ and $\kappa_4/\kappa_2=2$. Probe detunings are (a): $d_2=-2$. (b): $d_2=-5$. For large detunings, three results are indistinguishable.}
	\end{figure}
	\vskip 10 pt
	\noindent Figures 3(a)-3(d) show comparisons of numerical solutions with approximate analytical solutions for different probe powers and detunings. When $\kappa_4/\kappa_2=1$ the analytical solution achieves excellent agreement with the numerical solution in wide ranges of probe power and detuning for $C_0=1.00\pm 0.03$. When $\kappa_4/\kappa_2=2$, $d_2=-2$ we obtain remarkably accurate agreements in the probe power range of $S_0=$2 to 10 for a single amplitude adjustment constant of $C_0=1.17\pm 0.01$. For a fixed probe power of $S_0=10$, excellent agreements are obtained for a range of probe detuning $d_2$ from $-2$ to $-10$ with $C_0=1.15\pm 0.02$ \cite{r21}. Notice that the approximate analytical solution (red) show a slightly broader lineshape than the numerical solution (black). This is precisely because the replacement slightly overestimates the power-broadened line width.
	
	\vskip 10pt
	\noindent
	It is quite remarkable that accurate results from Eq. (6) can be obtained in such a tight range of amplitude adjustment constant $C_0$ for broad ranges of laser power and detuning. Indeed, with laser power varying from $S_0=$2 to 10 and in the probe detuning range of $d_2=-2$ to $-10$, Eq. (6) yields accurate results that are all within $\pm 5\%$ of that by the full numerical calculations. Figure 4 compares the numerical solution [Eqs. (4a) and (4b), solid black], analytical solution (Eq. (6), dashed red) and the simplified solution (Eq. (8), dash-dotted blue). For small probe detuning, which is the necessary operation condition for a single-probe scheme reported in \cite{r7}, the simplified solution without power broadening does not agree with neither numerical solution nor the approximated analytical solution [Fig. (4a)]. However, when the probe detuning increases, all three methods approach the same result. We note that even though a small probe detuning \cite{r7} enhances NMORE signal amplitude, the power broadened lineshape leads to a significant reduction in magnetic field detection sensitivity (i.e., much smaller slope near zero field). With a large probe detuning, the detection sensitivity is preserved but the NMORE signal amplitude is reduced significantly [Fig. 4(b)]. The recently reported colliding-probe bi-atomic magnetometer experiments and theory \cite{r15a,r15} overcome these issues, exhibiting excellent NMORE SNR, as well as increased field detection sensitivity at body temperature.  
	
	\vskip 10pt
	\noindent In conclusion, we have obtained the most accurate approximated analytical solutions to date for NMOREs of three and four-state systems with both probe absorption and power broadening. When the probe power broadening is neglected, we recover the known single-probe three and four-state system NMOREs. These general analytical solutions allow one to analyze multi-state magneto-optical rotation processes with excellent accuracy, revealing detailed effects and impacts of laser detuning, atomic hyper-fine splitting, as well as the probe power broadening on NMORE signal strength and magnetic field detection sensitivity.
	\vskip 10pt
	
	\noindent{Acknowledgments}
	Claire Deng thanks Dr. Changfeng Fang (SDU) for technical assistance on MATLAB coding. LD acknowledges the financial support from SDU.
	\vskip 10pt
	



\begin{thebibliography}{99}
\bibitem{r17} This is because frequency shifts cancels out in the calculations of NMORE which is the difference between the phase of two probe components. We verified this by comparing numerical calculations with and without $\beta_0$-dependent terms.
\bibitem{r18a} In fact, most if not all warm vapor single-probe magnetometers fall in this category.
\bibitem{r18} For a 10-cm cell the probe field change is limited to only a few percent due to the single-probe-based NMORE blockade.
\bibitem{r19} For warm vapor experiments such as reported in Ref.\cite{r7} with sophisticated 5-layer magnetic shielding and RF modulation technique the laser is typically detuning by about $\delta_p$=500 MHz (only about 1.5 Doppler linewidth) from $F'=1$ states. Since Rubidium hyper-fine splitting is $\Delta_0$=820 MHz, which is about 2 Doppler linewidths, we therefore chose $\Delta_0$=2 in unit of $\Gamma$.
\bibitem{r21} For instance, we first choose $C_0=1.15$ in Eq. (6) to obtained excellent agreement with numerical calculations for $d_2=-5$. Then, with a very narrow range of $C_0=1.15\pm 0.02$ excellent agreements with numerical calculations can be obtained for the entire detuning range from $d_2=-2$ to $-15$.
\end{thebibliography}
\end{document}